\documentclass[a4paper,12pt]{article}

\begin{document}
\topmargin 0pt \oddsidemargin 0mm

\renewcommand{\thefootnote}{\fnsymbol{footnote}}

\begin{titlepage}

\vspace{5mm}
\begin{center}
{\Large \bf Corrected Entropy-Area Relation and \\
Modified Friedmann Equations} \vspace{32mm}

{\large Rong-Gen Cai$^{a,}$\footnote{e-mail address:
cairg@itp.ac.cn}, Li-Ming Cao$^{a,}$\footnote{e-mail address:
caolm@itp.ac.cn}, Ya-Peng Hu$^{a,b,}$\footnote{e-mail address:
yapenghu@itp.ac.cn}}

\vspace{10mm} {\em $^a$ Institute of Theoretical Physics, Chinese
Academy of Sciences,\\
 P.O. Box 2735, Beijing 100190, China \\
 $^{b}$
  Graduate School of the Chinese Academy of Sciences, Beijing 100039, China}

\end{center}

\vspace{10mm} \centerline{{\bf{Abstract}}}
 \vspace{5mm}
Applying Clausius relation, $\delta Q=TdS$, to apparent horizon of a
FRW universe with any spatial curvature, and assuming that the
apparent horizon has temperature $T=1/(2\pi \tilde {r}_A)$, and a
quantum corrected entropy-area relation, $S=A/4G +\alpha \ln A/4G$,
where $\tilde {r}_A$ and $A$ are the apparent horizon radius and
area, respectively, and $\alpha$ is a dimensionless constant, we
derive modified Friedmann equations, which does not contain a bounce
solution. On the other hand, loop quantum cosmology leads to a
modified Friedmann equation $H^2 =\frac{8\pi G}{3}\rho
(1-\rho/\rho_{\rm crit})$. We obtain an entropy expression of
apparent horizon of FRW universe described by the modified Friedmann
equation. In the limit of large horizon area, resulting entropy
expression gives the above corrected entropy-area relation, however,
the prefactor $\alpha$ in the logarithmic term is positive, which
seems not consistent with most of results in the literature that
quantum geometry leads to a negative contribution to the area
formula of black hole entropy.

\end{titlepage}

\newpage
\renewcommand{\thefootnote}{\arabic{footnote}}
\setcounter{footnote}{0} \setcounter{page}{2}

\section{Introduction}

Due to the seminal work by Hawking~\cite{Haw}, nowadays it is
widely accepted that black hole behaves like a black body,
emitting thermal radiation, with a temperature proportional to its
surface gravity at the black hole horizon and with an entropy
proportional to its horizon area \cite{Haw,Bek}. The Hawking
temperature and horizon entropy together with the black hole mass
obey the first law of black hole thermodynamics~\cite{firstlaw}.
Hawking radiation is a quantum phenomenon. Therefore quantum
theory, gravitational theory and statistical physics are connected
each other in black hole thermodynamics. Einstein field equation
holds in a way that the Hawking process satisfies the first law of
thermodynamics. The development of black hole quantum physics and
seeking for a self-consistent quantum theory of gravity lead
people to consider the connection between the Einstein field
equation and the first law of thermodynamics, although at the
first glance, gravitational theory and thermodynamics have nothing
to do with each other, because they belong to two completely
different branches of science.

Assuming there is a proportionality between entropy and horizon
area, Jacobson~\cite{Jac} derived the Einstein field equation by
using the fundamental Clausius relation, $\delta Q= TdS$, connecting
heat, temperature and entropy. The key idea is to demand that this
relation holds for all the local Rindler causal horizon  through
each spacetime point, with $\delta Q$ and $T$ interpreted as the
energy flux and Unruh temperature seen by an accelerated observer
just inside the horizon. In this way, Einstein field equation is
nothing, but an equation of state of spacetime. Applying this idea
to $f(R)$ theory~\cite{Jac1,AC1,AC} and scalar-tensor
theory~\cite{AC1,CC1}, it turns out that a nonequilibrium
thermodynamic setup has to be employed. For another viewpoint, see
\cite{ES,WuYZ}.

In the spirit of Jacobson's derivation of Einstein field equation,
one is able to derive Friedmann equations of a
Friedmann-Robertson-Walker (FRW) universe with any spatial curvature
by applying the Clausius relation to apparent horizon of the FRW
universe~\cite{CK}. This works not only in Einstein gravitational
theory, but also in Gauss-Bonnet and Lovelock gravity theories. Here
a key ingredient is to replace the entropy area formula in Einstein
theory by using entropy expressions of black hole horizon in those
higher order curvature theories. For related discussions see also
\cite{FK,Dan,Bousso}. These results should closely relate to the
fact that Einstein field equation can be rewritten as an unified
first law~\cite{Hayward, Hayward1, Hayward2, Hayward3}. Indeed, at
the apparent horizon of FRW universe, the first Friedmann equation
can be cast to a form~\cite{CC1,AC2}, $dE_m=TdS +W_mdV$. Here
$E_m=\rho V$ is the total energy inside of the apparent horizon with
volume $V$, $W_m=(\rho-p)/2$ is the work density, $T$ and $S$ can be
regarded as the temperature and entropy associated with the apparent
horizon like a black hole horizon. This first law form also holds in
RSII bran world scenario, warped DGP model and even more complicated
case with a Gauss-Bonnet term in bulk~\cite{CC2,SWC1,SWC2}. Based on
this, one is able to find the relation between entropy expression of
apparent horizon and horizon geometry in bran world scenarios. These
results have been summarized in~\cite{Cai}. For further discussions
in this direction see~\cite{Ge,GW,WuYZ1,WWY}. On the other hand,
there also exist some studies in the relation between Einstein field
equation and first law of thermodynamics in the setup of black hole
spacetime~\cite{Pap}.

It is interesting to note that the Friedmann equations can be
derived by using Clausius relation to the apparent horizon of FRW
universe, in which entropy is assumed to be proportional to its
horizon area. Also it is well-known that the so-called area
formula of black hole entropy holds only in Einstein gravity. When
some higher order curvature term appears in some gravity theory,
the area formula has to be modified~\cite{Wald}. But it is clear
from Wald formula that horizon entropy of black hole must be a
function of horizon geometry. In this sense, it would be of great
interest to see whether one is able to derive modified Friedmann
equations if the entropy-area relation gets corrections by some
reason. For example, a logarithmic term often occurs in literature
\begin{equation}
\label{eq1}
 S= \frac{A}{4G} +\alpha \ln \frac{A}{4G},
\end{equation}
where $A$ is the horizon area and $\alpha$ is a dimensionless
constant. Such a term appears in studying black hole entropy in loop
quantum gravity (quantum
geometry)~\cite{Rov,Ash,Kau,Ghosh,Doma,Mei,Hod,Med,Chat1}, or in
discussing the correction to black hole entropy due to thermal
equilibrium fluctuation or quantum
fluctuation~\cite{Sol,Kas,Das,Chat1,Gour,Chat2}. However, even
within the quantum geometry, the value of the prefactor $\alpha$ is
in debate. Some reference gives $\alpha=-3/2$, for example,
\cite{Kau}; some works lead to $\alpha =-1/2$, for example,
\cite{Ghosh,Doma,Mei}; the paper~\cite{Hod} argued that $\alpha$
should be a positive integer and on the other hand, the author of
\cite{Med} argued that $\alpha$ should be equal to zero.

Applying the techniques of loop quantum gravity to homologous and
isotropic spacetime leads to the so-called loop quantum cosmology.
Due to quantum correction, the Friedmann equations get modified.
The big bang singularity is resolved and replaced by a quantum
bounce~\cite{LQC}. For a brief summary on loop quantum cosmology,
see~\cite{Ash2}.  Considering quantum correction, the modified
Friedmann equation turns out to be (in the case of
$k=0$)~\cite{Ash2}
\begin{equation}
\label{eq2} H^2=\frac{8\pi G}{3}\rho \left( 1-\frac{\rho}{\rho_{\rm
crit}}\right),
\end{equation}
where $\rho_{\rm crit}= \sqrt{3}/(32\pi G^2 \gamma^3)$, $\gamma$
is the so-called Barbero-Immirzi parameter.  This parameter could
be fixed as $0.2375$ in order to give the area formula of black
hole entropy in loop quantum gravity~\cite{Ash}. Due to the
corrected term in (\ref{eq2}), the big bang singularity is
replaced by a quantum bounce happening at $\rho=\rho_{\rm crit}$.

The aim of this paper is twofold. The first is to derive modified
Friedmann equations by applying Clausius relation to the apparent
horizon of FRW universe and assuming the horizon has an entropy
expression like (\ref{eq1}) or a more general form. The other is to
see whether the corrected entropy-area relation from loop quantum
gravity will lead to the modified Friedmann equation (\ref{eq2}) in
loop quantum cosmology because the approaches to reach these two
results seem different, although both of them are in the field of
loop quantum gravity.

The organization of this paper is as follows. In the next section we
obtain the modified Friedmann equations starting from the corrected
entropy-area relation (\ref{eq1}) by use of Clausius relation to the
apparent horizon of FRW universe. In Sec.~3 we get the entropy
expression associated with apparent horizon in FRW universe
described by the modified Friedmann equation (\ref{eq2}) in loop
quantum cosmology. The conclusion is included in Sec.~4.

\section{From corrected entropy-area relation to modified Friedmann
equations}

 To make the paper be self-contained, let us start the
case~\cite{CK} without the corrected term in the entropy-area
relation (\ref{eq1}). The FRW universe is described by metric
\begin{equation}
\label{eq3}
ds^{2}=-dt^{2}+a^{2}\left(\frac{dr^{2}}{1-kr^{2}}+r^{2}d\Omega
_{2}^{2}\right)=h_{ab}dx^{a}dx^{b}+\tilde{r}^2d\Omega _{2}^{2},
\end{equation}
where $x^0=t$, $x^1=r$, $h_{ab}= {\rm diag} (-1, a^2/(1-kr^2)$,
$\tilde{r}=a(t)r$ and $k$ denotes the spatial curvature. The
dynamical apparent horizon, a marginally trapped surface with
vanishing expansion, is determined by the relation $h^{ab}\partial
_{a}\tilde{r}\partial _{b}\tilde{r}=0$.  With this, it is easy to
find out the radius of the apparent horizon
\begin{equation}
\label{eq4}
 \tilde{r}_{A}=\frac{1}{\sqrt{H^{2}+k/a^{2}}},
\end{equation}
where $H=\dot a /a$ is the Hubble parameter and overdot stands for
the derivative with respect to cosmic time $t$.

Suppose that the energy-momentum tensor $T_{\mu \nu }$ of the matter
in the universe has the form of a perfect fluid$\ T_{\mu \nu }=(\rho
+p)U_{\mu }U_{\nu }+pg_{\mu \nu }$, where $\rho $ and $p$ are the
energy density and pressure, respectively. The energy conservation
law then leads to the continuity equation
\begin{equation}
\label{eq5} \dot{\rho}+3H(\rho +p)=0.
\end{equation}
Following \cite{Hayward2}, we define the work density $W$ and
energy-supply vector $\Psi$ as
\begin{equation}
\label{eq6} W=-\frac{1}{2}T^{ab}h_{ab},\ \ \Psi
_{a}=T_{a}^{b}\partial _{b}\tilde{r}+W\partial _{a}\tilde{r},
\end{equation}
where $T_{ab}$ is the projection of the $(3+1)$-dimensional
energy-momentum tensor $T_{\mu\nu}$ of matter in the FRW universe in
the normal direction of $2$-sphere. In our case, these are
\begin{equation}
W=\frac{1}{2}(\rho -p),\ \ \Psi _{a}=-\frac{1}{2}(\rho
+p)H\tilde{r}dt+\frac{1}{2}(\rho +p)adr.
\end{equation}
With this, we can compute the amount of energy crossing the apparent
horizon during the time internal $dt$~\cite{CK}
\begin{equation}
\label{eq8}
 \delta Q = -A\Psi =A(\rho +p)H\tilde{r}_{A}dt,
\end{equation}
where $A=4\pi \tilde{r}_A^2$ is the area of the apparent horizon. To
proceed, we make two assumptions: one is that the apparent horizon
has an horizon area entropy like black hole horizon; the other is
that the apparent horizon has a temperature, they have following
forms~\cite{CK}
\begin{equation}
\label{eq9}
 S=\frac{A}{4G}, \ \ T=\frac{1}{2\pi \tilde{r}_{A}}.
\end{equation}
Then using the Clausius relation $\delta Q=TdS$, we can reach
\begin{equation}
\label{eq10} \dot{H}-\frac{k}{a^{2}}=-4\pi G(\rho +p).
\end{equation}
Note that here we have used the relation
\begin{equation}
\label{eq11} \dot{\tilde{r}}_{A}=-H\tilde{r}_{A}^{3}(\dot
{H}-\frac{k}{a^{2}}),
\end{equation}
which comes directly from (\ref{eq4}). Furthermore, using the
continuity equation (\ref{eq5}) and integrating (\ref{eq10}), we can
obtain
\begin{equation}
\label{eq12} H^{2}+\frac{k}{a^{2}}=\frac{8\pi G}{3}\rho,
\end{equation}
where an integration constant, which is just the cosmological
constant, has been absorbed into the energy density $\rho$. Eqs.
(\ref{eq10}) and (\ref{eq12}) are nothing, but the Friedmann
equations of FRW universe. They are, of course, the concrete forms
of Einstein field equations, $R_{\mu\nu}-\frac{1}{2} g_{\mu\nu}R
=8\pi G T_{\mu\nu}$, in the FRW metric (\ref{eq3}).

Thus, we derive the Friedmann equations of a FRW universe with any
spatial curvature by applying the Clausius relation to apparent
horizon of the FRW universe and assuming that the apparent horizon
has an entropy satisfying the area formula like black hole horizon
and a temperature $T=1/2\pi \tilde r_{A}$.

Now we will apply the same idea to derive corresponding modified
Friedmann equations to the corrected entropy-area relation
(\ref{eq1}), although we did not yet know the modified Einstein
field equations due to quantum correction, which produces the
corrected entropy-area relation (\ref{eq1}).  To go on, there are
two key points, where seem worth mentioning here. Since we are still
considering a FRW universe, the assumption keeps unchanged, that the
apparent horizon has a temperature $T=1/(2\pi \tilde {r}_A)$,
because we know that Hawking temperature of black hole is completely
determined by spacetime metric (geometry), independently of gravity
theories, in which the black hole solution exists, while black hole
horizon entropy depends on gravity theories under consideration. The
other point is that since we are considering the perfect fluid
matter as source in the universe, in that case, the amount of energy
crossing the apparent horizon during the time internal $dt$ is still
given by (\ref{eq8}). That is to say, in this case, the only
difference is to replace area entropy $S=A/4G$ by the corrected
entropy-area relation (\ref{eq1}).

Thus the Clausius relation, $\delta Q=TdS$, in this time leads to
\begin{equation}
\label{eq13}
 A(\rho +p)H\tilde{r}_{A}dt=\frac{1}{2\pi
\tilde{r}_{A}}(\frac{1}{4G}+\frac{\alpha }{A})dA
\end{equation}
With the help of (\ref{eq11}), the above equation can be changed
into
\begin{equation}
\label{eq14} \left(1+\frac{4G\alpha}{A}\right) (\dot H
-\frac{k}{a^2}) = -4\pi G (\rho+p).
\end{equation}
Using the continuity equation, we can further rewrite (\ref{eq14})
as
\begin{equation}
\label{eq15}
 \left(1+\frac{\alpha G}{\pi} (H^2+\frac{k}{a^2})\right)
\frac{d (H^2+k/a^2)}{dt}= \frac{8\pi G}{3} \dot \rho.
\end{equation}
Integrating (\ref{eq15}) yields
\begin{equation}
\label{eq16}
 H^2 +\frac{k}{a^2} +\frac{\alpha G}{2 \pi} (H^2+\frac{k}
{a^2})^2 =\frac{8\pi G}{3}\rho,
\end{equation}
where an integration constant has been absorbed into the energy
density, again. Eqs.~(\ref{eq14}) and (\ref{eq16}) are nothing, but
the corresponding modified Friedmann equations to the corrected
entropy-area relation (\ref{eq1}).

Note that (\ref{eq16}) can be further rewritten as
\begin{equation}
\label{eq17}
 H^2+\frac{k}{a^2}=\frac{\pi}{\alpha G}\left(-1+\sqrt{1+\frac{16
 \alpha G^2}{3}\rho}\right).
 \end{equation}
 If $\alpha$ is viewed as a small quantity, expanding the right hand
 side of (\ref{eq17}), up to the linear order of $\alpha$, we have
 \begin{equation}
 \label{eq18}
 H^2+\frac{k}{a^2} \approx \frac{8\pi G}{3}\rho \left(1-\frac{4\alpha
 G^2}{3}\rho \right).
 \end{equation}
 Let us mention here that Eq.~(\ref{eq16})
 has another positive root for $(H^2+\frac{k}{a^2})$, but that
 solution has no classical limit as $\alpha \to 0$. We note that
 (\ref{eq18}) is quite similar to the modified Friedmann equation
 (\ref{eq2}) in loop quantum cosmology.  We see from
 (\ref{eq18}) that if $\alpha >0$, there seemly exists a bounce at
 \begin{equation}
\rho= \tilde {\rho}_{\rm crit}=\frac{3}{4\alpha G^2}.
\end{equation}
but this is not true. It is because (\ref{eq18}) is an approximate
solution, it holds only as $\rho/\tilde{\rho}_{\rm crit} \ll 1$. In
fact, it is easy to see from (\ref{eq16}) that there does not exist
any bounce solution in (\ref{eq16}). Notice the fact that
Eq.~(\ref{eq1}) is obtained in the limit of large horizon area, it
is not surprised that the corrected entropy-area relation does not
lead to a modified Friedmann equation with a bounce solution because
the latter is a nonperturbative effect.

When $\alpha<0$,  it is easy to see from (\ref{eq16}) that in this
case, there exists a de Sitter solution as the late-time attractor
of (\ref{eq16}): $H^2+k/a^2= 2\pi/(|\alpha|G)$.

In fact, the above procedure to derive the modified Friedmann
equations by using the corrected entropy-area relation can be
further generalized. For example, the next quantum correction to
black hole entropy (\ref{eq1}) is widely believed to have the form
$4G/A$, which means~\cite{Zhang}
\begin{equation}
S= \frac{A}{4G} + \alpha \ln \frac{A}{4G}+\beta \frac{4G}{A},
\end{equation}
where $\beta$ is another dimensionless constant. By using the same
procedure,  it is easy to show that in this case, the modified
Friedmann equations are
\begin{eqnarray}
\label{eq21} && \left(1+\alpha \frac{4G}{A} - \beta \frac{16
G^2}{A^2}\right) (\dot H -\frac{k}{a^2}) = -4\pi G (\rho+p),
 \\
 \label{eq22}
&& H^2 +\frac{k}{a^2} +\frac{\alpha
G}{2\pi}\left(H^2+\frac{k}{a^2}\right)^2 -\frac{\beta
G^2}{3\pi^2}\left(H^2+\frac{k}{a^2}\right)^3=\frac{8\pi G}{3}\rho,
 \end{eqnarray}
 where $A=4\pi/(H^2+k/a^2) $ is the area of apparent horizon.
Finally we stress that the above procedure also works well for a
more general case: the entropy is a function of horizon geometry.
For example, suppose the apparent horizon has an entropy with form
$f(A/4G)$, that is,
\begin{equation}
\label{eq23}
 S=f(x),\ \ x=\frac{A}{4G},
\end{equation}
where $f$ is an arbitrary function of horizon area. Applying the
Clausius relation $\delta Q=TdS$ to the apparent horizon of FRW
universe, we can obtain the second Friedmann equation concerning
the derivative  of Hubble parameter
\begin{equation}
\label{eq24}
 \left(\dot H -\frac{k}{a^2}\right) f'(x)=-4\pi G (\rho+p),
\end{equation}
where a prime stands for the derivative with respect to $x$. Using
the continuity equation, we can reach
\begin{equation}
\label{eq25}
 \frac{8\pi G}{3} \rho =-\frac{\pi}{G}\int \frac{f'}{x^2}d x.
 \end{equation}
 This is nothing, but the first Friedmann equation corresponding
 to the apparent horizon having the entropy form (\ref{eq23}).

\section{From modified Friedmann equations to corrected entropy-area
relation}

In this section, we will derive an entropy expression associated
with apparent horizon of a FRW universe described by the modified
Friedmann equation (\ref{eq2}) by using the method proposed in
\cite{CC2}. The starting point is the unified first
law~\cite{Hayward,Hayward1,Hayward2}: The Einstein field equation
can be rewritten as
\begin{equation}
\label{eq26}
 dE =A \Psi +W dV,
 \end{equation}
 where $A= 4\pi \tilde {r}^2$ and $V= \frac{4\pi}{3} \tilde {r}^3$ are area and volume
 of $3$-dimensional sphere with radius $\tilde {r}$, $E$ is the Misner-Sharp energy
\begin{equation}
 E= \frac{\tilde {r}}{2G} (1-h^{ab}\partial_a \tilde r \partial_b
 \tilde r),
 \end{equation}
 $\Psi$ and $W$ are defined by (\ref{eq6}). According to
 thermodynamics, entropy is associated with heat flow as $\delta
 Q=TdS$, and heat flow is related to change of energy of a given
 system. As a consequence, entropy is finally associated with the
 energy-supply term. The latter can be rewritten as
 \begin{equation}
 \label{eq28}
 A \Psi= \frac{\kappa}{8\pi G} dA +\tilde {r} d(E/\tilde{r}),
 \end{equation}
 where $\kappa$ is the surface gravity defined as
\begin{equation}
\kappa= \frac{1}{2\sqrt{-h}}\partial_a(\sqrt{-h}h^{ab}\partial_b
\tilde{r}).
\end{equation}
On the apparent horizon, the last term in (\ref{eq28}) vanishes,
then one can assign an entropy $S=A/4G=\pi \tilde {r}_A^2/G$ to the
apparent horizon with radius $\tilde{r}_A$. That is, in Einstein
gravity, apparent horizon can be assigned an entropy proportional to
its area like black hole entropy. After projecting along the vector
$\xi = \partial_t -(1-2\epsilon)Hr\partial_r$ with $\epsilon=
\dot{\tilde{r}}_A/2H\tilde{r}_A$, we can get the first law of
thermodynamics of apparent horizon~\cite{CC2}
\begin{equation}
\label{eq30}
 \langle dE,\xi\rangle =\frac{\kappa}{8\pi G}\langle dA,\xi\rangle
 +\langle WdV,\xi\rangle,
 \end{equation}
 where $\kappa= -(1-\epsilon)/\tilde{r}_A$ by definition. Next we
 can obtain the entropy expression associated with apparent horizon
 by using Clausius relation, $\delta Q=TdS$, here the heat flow
 $\delta Q$ should be given by pure matter energy-supply $A\Psi_m$.
 To get pure matter energy-supply vector $\Psi_m$, we could rewrite
 (\ref{eq2}) as
 \begin{equation}
 \label{eq31}
 H^2 +\frac{k}{a^2} = \frac{8\pi G}{3} (\rho +\rho_e),
 \end{equation}
 where we have added a curvature term $k/a^2$, and $\rho_e
 =-\rho^2/\rho_{\rm crit}$. Note that whatever the value of $k$ is, the following
 conclusion is unchanged. Using continuity equation, we can obtain
 the effective pressure corresponding to the effective energy
 density, $p_e = -\rho (\rho+2p)/\rho_{\rm crit}$. Then we have
 associated energy-supply vector $\Psi_e$ and work density $W_e$
 \begin{eqnarray}
 \label{eq32}
 && \Psi_e = \frac{\rho}{\rho_{\rm crit}}(\rho+p) H\tilde {r} dt
 -\frac{\rho}{\rho_{\rm crit}}(\rho+p) a dr,
 \nonumber \\
 && W_e= \frac{\rho }{\rho_{\rm crit}}p.
 \end{eqnarray}
 We have from (\ref{eq30}) that the heat flow of pure matter
 \begin{eqnarray}
 \label{eq33}
&&  \delta Q \equiv \langle A\Psi_m, \xi \rangle =
  \frac{\kappa}{8\pi G} \langle dA,\xi \rangle-\langle A\Psi_e,\xi
  \rangle
   \nonumber \\
  &&~~~~~ =-\frac{H A \epsilon (1-\epsilon)}{2\pi G}
  \frac{1}{\sqrt{\tilde{r}_A^2-\frac{3}{2\pi G \rho_{\rm crit}}}}
  \nonumber \\
  &&~~~~= T \langle \frac{2\pi \tilde{r}_A^2}{G} \frac{d\tilde {r}_A}{\sqrt{\tilde{r}_A^2-\frac{3}{2\pi G
  \rho_{\rm crit}}}},\xi \rangle
  \nonumber \\
  &&~~~~= T \langle dS,\xi \rangle
  \end{eqnarray}
  where we have defined $T=\kappa/2\pi$. Integrating (\ref{eq33}),
  we reach
  \begin{equation}
  \label{eq34}
  S= \frac{\pi \tilde{r}_A}{G}\sqrt{\tilde{r}_A^2-\frac{3}{2\pi
  G\rho_{\rm crit}}} + \frac{3}{2\pi G^2\rho_{\rm crit}}\ln (\tilde
  {r}_A +\sqrt{\tilde{r}_A^2-\frac{3}{2\pi
  G\rho_{\rm crit}}}) +C,
  \end{equation}
  where $C$ is a constant, whose value should be determined by some
  physical condition. It is easy to show that with the entropy
  (\ref{eq34}), the first law of thermodynamics for apparent horizon
  holds~\cite{AC2}
  \begin{equation}
  dE_m= TdS +W_m dV,
  \end{equation}
  where $E_m=\rho V$, $W_m= (\rho-p)/2$ and $V=4\pi \tilde{r}_A^3/3$ are
total energy inside the apparent horizon, work density and volume of
the apparent horizon, respectively. In addition, let us notice that
in the limit of large apparent horizon, entropy (\ref{eq34}) has the
form
\begin{equation}
\label{eq36}
 S = \frac{A}{4G} + \frac{3}{4\pi G^2 \rho_{\rm crit}} \ln \frac{A}{4G}
  + o (\frac{1}{A}) + C_0.
  \end{equation}
  where $A=4\pi \tilde{r}_A^2$ is the area of apparent horizon,
  $C_0$ is another constant and $3/4\pi G^2 \rho_{\rm crit}=8 \sqrt{3}\gamma^3 \approx
  0.1856$. In this case, entropy expression (\ref{eq36}) has the
  same form as (\ref{eq1}). But here the prefactor in the logarithmic term $\alpha'= 3/4\pi G^2 \rho_{\rm crit}$ is
   positive. Note that  (\ref{eq1}) is  given in
  the limit of large horizon area. Therefore although the form of our result
  (\ref{eq36}) is the same as (\ref{eq1}), the corrected term gives an opposite contribution
  to the area entropy as the one quantum geometry does.

 \section{Conclusions}
 In summary, applying the Clausius relation $\delta Q=TdS$ to apparent
 horizon of a FRW universe with any spatial curvature and assuming
 that the apparent horizon has  temperature $T=1/(2\pi \tilde
 {r}_A)$, and an entropy obeying the well-known one quarter horizon
 area formula like black hole horizon, one is able to derive
 Friedmann equations governing the dynamical evolution of the
 universe. We followed the same idea to derive modified Friedmann
 equations from a quantum corrected entropy-area relation
 (\ref{eq1}). However, resulting modified Friedmann
 equations do not contain a bounce solution. This result is
 not in contradiction with (\ref{eq2}) since (\ref{eq1}) holds
 only in the limit of large horizon area.  We also derived
 corresponding modified Friedmann equations in the case with the
 apparent horizon having an entropy which is an arbitrary function
 of  horizon area.

On the other hand, starting from the modified Friedmann equation
(\ref{eq2}) in loop quantum cosmology, we obtained an entropy
expression associated with the apparent horizon of a FRW universe by
use of the unified first law for dynamic horizon. In the limit of
large horizon radius, resulting entropy indeed has the same form as
the corrected entropy-area relation (\ref{eq1}). However, the
prefactor in the logarithmic term is positive, which seems in
contradiction with most of results in literature that quantum
geometry gives a negative contribution to the area formula of black
hole entropy.

 Finally, let us mention that although we
 derived modified Friedmann equations corresponding to the
 corrected entropy-area relation (\ref{eq1}) by using Clausius
 relation, it would be of great interest to see whether one is able to
  get modified Einstein field equation by following
  Jacobson~\cite{Jac}. If this works, it further shows
  that given a thermodynamical relation between entropy and
  geometry, one is able to derive corresponding modified Einstein
  field equation, showing an interesting connection between them.


\section*{Acknowledgments}
RGC thanks Y.G. Ma for helpful discussions on loop quantum gravity
and loop quantum cosmology.  This work was supported partially by
grants from NSFC, China (No. 10325525 and No. 90403029), and a grant
from the Chinese Academy of Sciences.

\end{document}